# SafeLoad: Efficient Admission Control Framework for Identifying Memory-Overloading Queries in Cloud Data Warehouses

Yifan Wu
Zhejiang University[†]
yifan.wu@zju.edu.cn

Yuhan Li
Zhenhua Wang
Alibaba Cloud Computing
{lyh200442,wzh420090}@alibaba-inc.com

Zhongle Xie
School of Software Technology, Zhejiang University
xiezl@zju.edu.cn

Dingyu Yang
Ke Chen
Lidan Shou
Zhejiang University[†]
{yangdingyu,ck,should}@zju.edu.cn

Bo Tang
Southern University of Science and Technology
ltangb3@sustech.edu.cn

Liang Lin
Alibaba Cloud Computing
yibo.ll@alibaba-inc.com

Huan Li[*]
Gang Chen
Zhejiang University[†]
{lihuan.cs,cg}@zju.edu.cn

## ABSTRACT

Memory overload is a common form of resource exhaustion in cloud data warehouses. When database queries fail due to memory overload, it not only wastes critical resources such as CPU time but also disrupts the execution of core business processes, as memory-overloading (MO) queries are typically part of complex workflows. If such queries are identified in advance and scheduled to memory-rich serverless clusters, it can prevent resource wastage and query execution failure. Therefore, cloud data warehouses desire an admission control framework with high prediction precision, interpretability, efficiency, and adaptability to effectively identify memory-overloading queries. However, existing admission control frameworks primarily focus on scenarios like SLA satisfaction and resource isolation, with limited precision in identifying MO queries. Moreover, there is a lack of publicly available MO-labeled datasets with workloads for training and benchmarking. To tackle these challenges, we propose SafeLoad, the first query admission control framework specifically designed to identify MO queries. Alongside, we release SafeBench, an open-source, industrial-scale benchmark for this task, which includes 150 million real queries. SafeLoad first filters out memory-safe queries using the interpretable discriminative rule. It then applies a hybrid architecture that integrates both a global model and cluster-level models, supplemented by a misprediction correction module to identify MO queries. Additionally, a self-tuning quota management mechanism dynamically adjusts prediction quotas per cluster to improve precision. Experimental results show that SafeLoad achieves state-of-the-art prediction performance with low online and offline time overhead. Specifically, SafeLoad improves precision by up to 66% over the best baseline and reduces wasted CPU time by up to 8.09× compared to scenarios without SafeLoad.

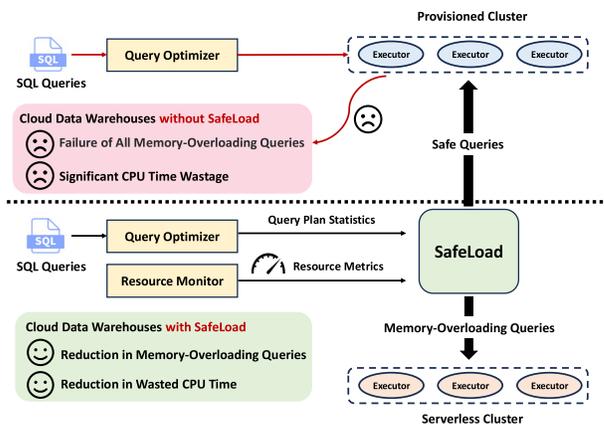

Figure 1: The role of SafeLoad in cloud data warehouses.

## 1 INTRODUCTION

Cloud data warehouses, such as Redshift [1], Snowflake [8], and AnalyticDB [30], have reached unprecedented scales, processing hundreds of billions of queries per day. These systems typically operate with thousands of database clusters under sustained high load, making resource contention the norm rather than the exception. In such environments, effective resource management is critical, as unregulated query execution can compromise system stability and performance. Query admission control [19, 27] plays a vital role in deciding whether a query should proceed to execution, with the overarching goal of preventing resource exhaustion and ensuring overall system efficiency.

**Memory overload**, characterized by query demand exceeding available capacity, remains a persistent challenge in cloud databases, frequently leading to performance degradation and execution failures. These incidents not only disrupt critical business workflows but also incur substantial resource wastage. Our empirical analysis of the collected data from an industrial data warehouse AnalyticDB [30] shows that execution failures stemming from memory overload result in a loss of up to 17 CPU hours per failure incident. Given the scale and impact of such query failures, an admission control framework that proactively identifies MO queries before execution would be highly beneficial for both cloud providers and users. This motivates the design of a dedicated admission control framework for MO query detection presented in this work. As illustrated in Figure 1, the framework, referred to as SafeLoad, identifies these MO queries in advance, redirecting them to memory-rich serverless clusters for safe execution. Without such a framework, MO queries are admitted into provisioned clusters, only to fail due to memory shortages, wasting the CPU time consumed up to the point of failure.

Drawing on industrial case studies of cloud data warehouses handling MO queries [6, 7], together with lessons learned from open-source systems [28], we identified three key goals that an ideal admission control framework for MO query detection should achieve:

- **Goal 1: High Prediction Precision**. The ideal solution should maximize its ability to accurately identify MO queries and redirect them to memory-rich serverless clusters. For cloud providers, this ensures system stability, while for users, it prevents query failures and conserves valuable resources.
- **Goal 2: Interpretability and Efficiency**. The ideal solution should provide clear explanations to users, enabling them to understand why specific queries are rejected. Furthermore, the prediction overhead should be minimal, ensuring that query latency remains unaffected. For cloud providers, implementing this solution should incur minimal operational overhead, making it feasible for widespread adoption [12].
- **Goal 3: Adaptability to Different Database Clusters**. Given the significant differences in query patterns across various database clusters, an ideal solution should adapt to the specific characteristics of each cluster. This adaptability enables the solution to better distinguish between non-MO and MO queries within a cluster, thereby reducing the number of mispredictions.

However, existing admission control approaches predominantly focus on query latency [19, 22, 24, 27], Service-Level Agreement (SLA) compliance [26], resource isolation [24], and throughput optimization [19]. They are designed for other scenarios and thus fail to directly address the challenge of identifying MO queries. The most relevant studies [19, 21, 29] are also limited to estimating the memory consumption of queries. Moreover, as reported in Section 6, the highest precision achieved by these methods is only 0.5225, which is too low to support practical deployment. Meanwhile, the goals of interpretability and adaptability remain unexplored.

Next, we discuss the technical challenges associated with achieving the three goals mentioned above. First, for **Goal 1**, due to the lack of in-depth feature analysis, it remains unclear which features are useful for identifying MO queries. Additionally, the lack of publicly available labeled datasets related to MO queries makes it challenging to construct a competent model. Furthermore, the scarcity of MO queries compared to non-MO queries further complicates the task of accurately identifying MO queries. Specifically, within the query set from the real production environment we analyzed, the number of non-MO queries can exceed that of MO queries by up to 20,276 times. Second, for **Goal 2**, it is necessary to explore efficient, human-interpretable heuristic rules for distinguishing MO and non-MO queries. Such heuristics can enhance users' understanding of the decision-making process within the admission control framework. On the other hand, while training machine learning (ML) models to capture the complex feature distributions of MO queries is essential, further investigation is needed to achieve efficient training and minimize prediction overhead. Third, for **Goal 3**, memory-overloading behavior varies significantly across database clusters, leading to substantial feature distribution shifts in some clusters. This heterogeneity makes it difficult for a global model trained on all data to achieve accurate predictions across all clusters (see Section 3).

To address these challenges, we propose SafeLoad, an efficient framework for MO query admission control. For **Goal 1**, we analyze features across seven categories from query and cluster perspectives. Using SHAP [14], KDE [16], and t-SNE [23], we quantify and visualize feature importance, distributions, and high-dimensional patterns, forming the basis for high-precision detection. To facilitate further research, we release SafeBench, an open-source benchmark with 150M production queries. We also design a hybrid architecture combining a global model with cluster-level models, enhanced by a misprediction correction module. Specifically, SafeLoad builds a FAISS-based vector index [9] for each cluster to store misclassified MO queries. New queries are matched against this index to correct mispredictions, significantly improving precision.

Regarding **Goal 2**, we propose a discriminative rule filtering module that efficiently eliminates the majority of non-MO queries using simple, human-readable rules. This approach not only reduces the computational overhead of learned models but also enhances interpretability by providing clear and actionable explanations for admission decisions. As for **Goal 3**, we propose a self-tuning quota management module that dynamically estimates a quota cost for each query by considering model prediction entropy and false negative count. The quota refers to the limit on the number of queries predicted as positive, which restricts the number of queries forwarded to serverless clusters for execution. Moreover, this module adapts the quota cost in real time based on the prediction behaviors observed on each database cluster. This module reduces mispredictions and minimizes the risk of true memory overloads.

We compare SafeLoad with existing approaches on the industrial-grade SafeBench benchmark. The results show that SafeLoad improves precision by up to 66% over the best baseline and reduces CPU time wasted by MO queries by as much as 8.09× compared to cases without SafeLoad. SafeLoad integrates well-engineered techniques into a novel and unified framework for MO query detection. Our ablation study (see Section 6.3) shows that every component is essential, and their integration achieves accuracy beyond what any part can deliver alone. We conducted a detailed ablation study to examine the contribution of each component, and the results demonstrate that the whole is greater than the sum of its parts. Each component plays an indispensable role in achieving the overall effectiveness of the framework.

The key contributions of this paper are summarized as follows:
- We introduce a novel research problem of identifying MO queries before query execution. This problem aligns with the real-world issues encountered by modern data warehouses (Section 2).
- We introduce SafeLoad, the first query admission control framework for identifying MO queries. The architecture of SafeLoad consists of three key components: discriminative rule filtering, predictive model with correction, and self-tuning quota management (Section 4).
- We release the first industrial-grade benchmark for studying MO queries, SafeBench, constructed from real query workloads and cluster-level information collected from AnalyticDB, which includes 150 million queries. SafeBench fills a critical gap in this research area, where publicly available benchmarks have been lacking (Section 5).
- We conduct an in-depth comparison between SafeLoad and existing methods on the industrial-grade benchmark SafeBench. Experimental results show that SafeLoad achieves state-of-the-art prediction performance with low prediction overhead. We also conduct ablation studies to confirm the necessity of each SafeLoad component (Section 6).

In addition, Section 3 presents feature analysis; Section 7 reviews related work; and Section 8 concludes with future directions.

## 2 PROBLEM STATEMENT

We start by introducing the definition and root causes of memory-overloading queries observed in production environments. Subsequently, we provide a formal problem definition and clarify the key metrics relevant to the business scenarios.

**Memory-Overloading Queries.** Memory overload, often referred to as *Out-of-Memory* (OOM) exceptions, is a critical issue in cloud database systems, often leading to query failures and system instability. These exceptions occur when a query's memory requirements exceed the available system resources during execution, resulting in operational disruptions. Formally, we define a *Memory-Overloading Query (MO Query)* as follows:

**Definition 1 (Memory-Overloading Query, MO Query).** *In cloud database systems, a memory-overloading query is defined as a database query whose memory demand at any point during its execution exceeds the available memory capacity allocated for the user's database cluster, resulting in a memory overload condition.*

MO queries can be classified into two types based on their causes and system-level implications: (1) *High Query Memory Demand*: This type arises when specific blocking operators, such as `Group By`, `Sort`, `Window`, and `Hash Join`, require substantial memory to materialize intermediate results. When their memory requests exceed the reserved pool, the query fails. Similarly, streaming operators like `TableScan` may trigger memory overload when incremental allocations across concurrent queries exhaust the shared memory pool. (2) *Cluster-Wide Memory Pressure*: This occurs when the overall memory across all nodes is insufficient to support all active queries. In such cases, the system may terminate high-memory-consuming queries to maintain stability, even if those queries have not individually exceeded their limits. These two memory overload types manifest differently in query plan and system-level metrics — queries with complex query operators are more prone to the first type, while high cluster-wide memory utilization increases the likelihood of the second.

**Table 1: Explanation of key terms, highlighting benefits and drawbacks from a real-world deployment perspective.**

| Term | Definition / Interpretation |
|---|---|
| True Positive (TP) | An MO query correctly predicted as positive; ***Cloud Provider***: Prevents database overload by offloading MO queries; ***User***: Ensures reliable execution of MO queries. |
| True Negative (TN) | A non-MO query correctly predicted as negative; ***Cloud Provider***: Conserves serverless resources and minimizes costs; ***User***: Avoids unnecessary offloading and execution delays. |
| False Positive (FP) | A non-MO query incorrectly predicted as positive; ***Cloud Provider***: Wastes serverless resources and inflates costs; ***User***: Causes delays and unnecessary quota consumption. |
| False Negative (FN) | An MO query incorrectly predicted as negative; ***Cloud Provider***: Risks database overload and degraded performance; ***User***: Causes failure in execution of MO queries and waste of resources. |
| Accuracy | $\frac{TP+TN}{TP+TN+FP+FN}$. High accuracy reflects overall model reliability in balancing resources and user needs. |
| Precision | $\frac{TP}{TP+FP}$. ***Cloud Provider***: Limits unnecessary offloading, improving resource efficiency; ***User***: Reduces disruptions for non-MO queries. |
| Recall | $\frac{TP}{TP+FN}$. ***Cloud Provider***: Ensures most MO queries are offloaded, avoiding overloads; ***User***: Maximizes handling of MO queries. |
| F1-score | $\frac{2 \times \text{Precision} \times \text{Recall}}{\text{Precision}+\text{Recall}}$. Balances precision and recall for optimal resource use and user satisfaction. |

**Problem Definition and Application Scenario.** Next, we introduce the research problem addressed in this study. The task of our work is to predict whether an SQL query will experience memory overload after being processed by the query optimizer, based on the query plan statistics and cluster resource metrics. In cloud databases, users execute SQL queries on provisioned clusters. Before execution, each query is analyzed to predict whether it might cause memory overload. Queries predicted to exceed memory limits are redirected to a serverless cluster with sufficient resources, which is usually maintained by the cloud provider for system stability. However, executing queries on the serverless cluster incurs additional quota costs, measured in CPU time.

To minimize costs, it is critical to improve prediction precision and reduce erroneous dispatches. Moreover, the cloud provider should implement mechanisms to limit the number of dispatched queries, preventing excessive resource consumption caused by mispredictions (see our discussions in Section 4.4).

Within this context, we formally define the problem as follows.

**Problem 1 (MO Query Detection).** *Let $s$ be an SQL query, with $P(s)$ representing its query plan statistics after optimization, and $R(c)$ denoting the resource metrics of cluster $c$ where $s$ is executed. We define MO query detection as a binary classification problem:*

$$f : \{P(s), R(c)\} \to \{0, 1\},$$

*where $f = 1$ indicates that the query is predicted to cause memory overload, and $f = 0$ indicates it is not.*

**Distinguishing MO Query Detection from Memory Estimation.** MO query detection and query memory consumption estimation are related but differ in objectives and typical methodology. The former is formulated as a binary classification problem:

decide whether a query, conditional on the current system state, will trigger a memory-overload failure; the latter is a regression problem that predicts peak memory consumption [19, 21, 29]. A baseline heuristic estimates a query's peak memory consumption and compares it with the available memory; however, our empirical evaluation (Section 6.2) indicates that this approach is substantially less accurate than SafeLoad. This discrepancy arises due to: 1) dynamic memory demand (transient spikes can induce overload events at any point, not necessarily at the predicted peak); and 2) time-varying availability (the free-memory budget can change between query submission and execution, rendering comparisons against the initial snapshot unreliable).

> **Key Performance Indicator**. Queries causing memory overloads are labeled as *positive*, while successful ones are *negative*. Table 1 introduces FP, FN, precision, recall, and F1-score in business contexts. In practice, excessive FPs misclassify non-MO queries, wasting user quota and raising costs, while excessive FNs overlook MO queries, straining provisioned clusters and disrupting workflows.
> From a business view, higher *precision* lowers user cost, while better *recall* improves resource use and query success. The *F1-score*, balancing both, offers a holistic measure. Thus, this study highlights **precision** and **F1-score** as the most **critical metrics** for deployment.

## 3 FEATURE ANALYSIS

To address the MO Query Detection problem, we first conduct an in-depth exploration using a naive predictor over real workloads, focusing on the three goals outlined in Section 1. Specifically, our goal is to quantitatively identify the key features that guide the prediction of MO queries, in order to improve precision (Goal 1), interpretability and efficiency (Goal 2), and adaptability (Goal 3).

**Setups and Basic Statistics**. We collect real production data from AnalyticDB [30], namely the query logs obtained from two representative datasets collected over continuous days. The first dataset comprises 854 clusters and 52 million queries, among which 4,014 are MO queries. The second dataset includes 862 clusters and 50 million queries, with 2,508 identified as MO queries. We employ the widely adopted XGBoost model as the naïve predictor [19, 21, 29]. The model is trained on the first dataset and subsequently evaluated on the second dataset to assess its generalization performance. The input features used for model training were derived from the query optimizer and resource monitor. The optimizer can provide query-level statistics, while resource monitors can expose hundreds of performance metrics at either cluster-level or hardware-level.

**Analysis for Goal 1**. The first goal is to improve the precision of the predictor, which needs to effectively select the indicative features of memory overload. Therefore, we conducted a feature importance analysis using Shapley Additive Explanations (SHAP) [14], which quantifies the contribution of each feature to the model's predictions. From this analysis, we categorized the key features into seven groups, as outlined in Observation 1. This taxonomy informs both our feature extraction process and the design of the learned model (cf. Sections 4.1 and 4.3).

**Observation 1.** *As shown in Table 2, SHAP analysis identifies four query-level feature groups — Operator Count, Operator Cardinality, Memory-Intensive Operators, and Execution Configuration — as critical for detecting MO queries, as they capture query complexity, data size, and execution behavior. At the cluster level, three feature groups — Resource Metrics, Cluster Configuration, and OOM Indicators — effectively reflect resource availability and failure trends, key factors in predicting memory overloads.*

**Analysis for Goal 2**. To ensure interpretability and efficiency, we could seek to determine whether there exists a handful of features that can effectively distinguish MO and non-MO queries. These features can be cited as a clue explaining why the query is predicted as an overload, as well as filter out the non-MO queries for efficiency requirements. We focus on analyzing the common features of the two types of MO queries mentioned in Section 2, including counts of `Join`, `Window`, `Aggregation`, and `Sort` nodes, the total output size of `TableScan` nodes, and whether clusters experienced OOM events previously, regarding their discriminative power in our analysis. To visualize, we applied Kernel Density Estimation (KDE) [16] on the features, and the results are as shown in Figure 2. In the figure, blue regions represent MO queries, while red regions represent non-MO queries. The $x$-axis denotes feature values, and the $y$-axis indicates probability density. Several clear patterns emerged, for example, most non-MO queries have `Join` node counts of 0 or 1, `TableScan` output sizes under 1 MB, and are associated with clusters that did not experience OOM events. Consequently, we have the following observation:

**Observation 2.** *Certain features, such as the number of `Join` nodes and the output size of `TableScan`, exhibit strong distinguishability for non-MO queries.*

Observation 2 reveals the potential of using simple, interpretable feature rules to efficiently filter out a large portion of non-MO queries, reducing the computational burden of MO query detection. This inspires the design of a discriminative rule filtering module before applying a learned classifier, as detailed in Section 4.2.

Moreover, we evaluate industry heuristics for the MO query detection task. Consistent with Observation 2, structural features — such as the size of `TableScan` nodes and the number of `Join` nodes — carry significant discriminatory signal between MO and non-MO queries. However, our empirical results (Section 6.2) show that these heuristics achieve very high overall accuracy (up to 0.99) yet extremely low precision on the MO class (lower than 0.0258); the resulting false positives impose non-trivial operational cost and limit their practicality in production settings. These findings motivate the SafeLoad framework's multi-layer design: heuristics serve as a first-pass filter to remove obvious non-MO queries, followed by lightweight ML classifiers that substantially increase precision and improve MO query detection reliability.

**Observation 3.** *While heuristic rules are fast and interpretable, they are insufficient for achieving high-precision MO query detection. ML models are required to further distinguish MO queries with ambiguous characteristics that heuristic rules cannot differentiate.*

**Analysis for Goal 3**. For the last goal, adaptability to different database (DB) clusters, we analyze the feature distributions of MO queries in both intra- and inter-cluster manner. Using the t-SNE

Table 2: A non-exhaustive feature set for identifying MO queries. Detailed descriptions are available in the SafeBench benchmark (Section 5).

| Category | Feature Group | Description | Count |
| --- | --- | --- | --- |
| Query-level | Operator Count | Number of distinct operators (23 types in total), such as the count of `join` nodes in the query. | 23 |
| | Operator Cardinality | Cardinality statistics for 13 operators, covering 8 metrics per operator, including total and maximum values of the output size and row count. These operators are selected based on error reports from MO queries. | 104 |
| | Memory-intensive Operators | Fine-grained features of operators typically associated with high memory usage (i.e., `join`, `aggregation`, `window`, `sort`); includes child node cardinality for `join` and total number of `varchar`-typed grouping keys in `aggregation`. | 19 |
| | Execution Configuration | Configuration information during query execution scheduling, including the total degree of query parallelism and an indicator of whether the SQL query explicitly specifies an execution mode (e.g., `batch` mode), which may impact memory behavior. | 2 |
| Cluster-level | Resource Metrics | Cluster resource usage metrics collected at 1-minute granularity, including QPS, average memory pool utilization, and CPU utilization. | 8 |
| | Cluster Configuration | Static configuration of the provisioned cluster, including the number of CPU cores and the resource group ID. | 6 |
| | OOM Indicator | Number of OOM events observed in the corresponding cluster on the previous day. | 1 |

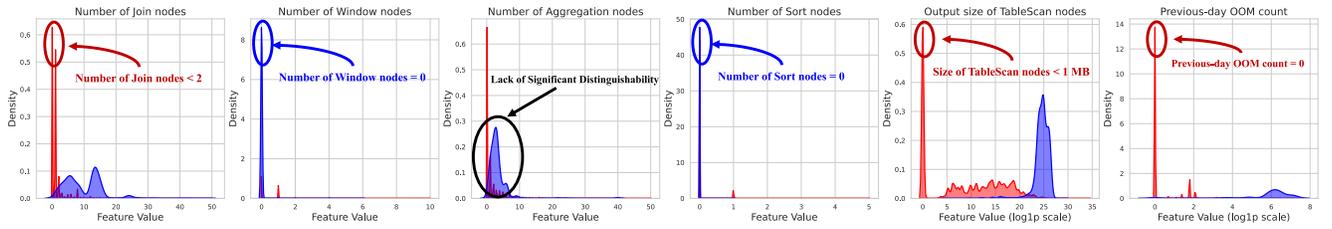

Figure 2: Kernel density estimation of selected features for distinguishing MO and non-MO queries. Blue regions correspond to MO queries, while red regions represent non-MO queries. A $\log(1 + x)$ transformation is applied to the last two subplots to handle large-scale features.

algorithm [23], we project high-dimensional MO query features into a 2D space for visualization. As shown in Figure 3, we analyze the top-5 DB clusters[*] with the largest MO query populations (labeled with pseudo cluster names A to E). The results reveal that MO queries tend to form numerous small clusters. Queries within the same cluster generally exhibit higher feature similarity compared to those in different clusters, although variability exists even within the same cluster. Therefore, we formulate the following observation:

**Observation 4.** *MO query features vary significantly across DB clusters. Additionally, each MO query is often accompanied by a small group of highly similar counterparts.*

Observation 4 highlights two key insights: (1) Significant feature differences across clusters call for both global and cluster-specific learned models. (2) MO queries often form small, highly similar groups, arising from users repeatedly issuing the same query after an initial memory-overloading failure in an attempt to retry the task. Furthermore, the strong feature similarity within query groups offers an opportunity to design a misprediction correction mechanism, where temporally neighboring queries can provide corrective signals. These ideas are further explored in Section 4.3.

Furthermore, how frequently do similar queries reappear in real-world workloads? This question is crucial for validating our misprediction correction mechanism, as high repetition rates enable effective learning from historical mispredictions. Our analysis reveals compelling seasonal patterns that strongly support our design. First, at the query level, 77.78% and 77.30% of queries share identical SQL text in the analyzed datasets. More notably, when analyzing query-level features, repetition rates reach 91.67% and 91.29%, indicating that different SQL texts often exhibit identical statistical patterns. Second, combining query-level and cluster configuration features, repetition rates remain substantial at 80.97% and 79.72%. These high frequencies of recurring query-configuration pairs demonstrate that our correction mechanism can effectively leverage historical misprediction patterns, providing strong empirical support for incorporating historical feedback in our design.

[*]The top-5 clusters are selected based on their highest MO query counts.

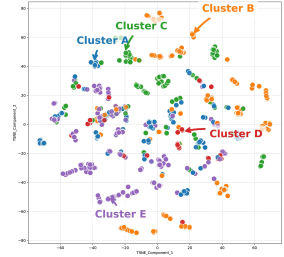

Figure 3: t-SNE visualization of MO query features from the top-5 clusters with the most MO queries.

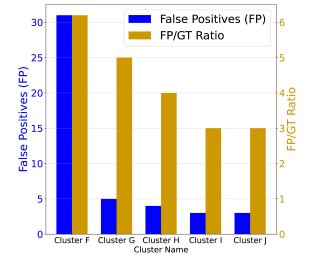

Figure 4: Analysis of misclassification behavior in the top-5 clusters with the highest FP/GT ratios.

Note that for inter-cluster adaptation, a critical challenge in MO query detection is the cost incurred by false positives. Hence, we further analyze the misclassification behavior across database clusters, specifically focusing on the top five clusters (pseudo-named F to J) with the highest False Positive (FP) to Ground Truth (GT) ratios, where model performance is notably poor. For instance, in Cluster F (Figure 4), the model correctly identifies only 5 true MO queries but incorrectly predicts 31 non-MO queries as MO. This results in an FP-to-GT ratio of 6.2, leading to 6.2 times higher quota consumption than necessary. Similarly, clusters G to J also display significantly more false positives than true MO queries, with varying degrees of severity. To gain insights, we introduce and examine the model's confidence scores, which indicate the likelihood of a query being classified as positive. On average, true positives have a confidence score of 0.99, false positives 0.91, false negatives 0.04, and true negatives 0. These scores reveal a clear distinction in confidence between correct predictions and mispredictions.

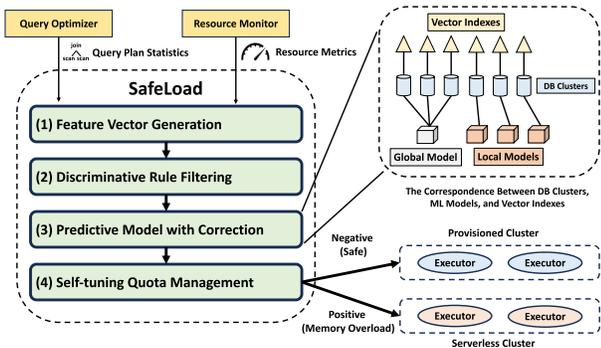

Figure 5: The overview of SafeLoad.

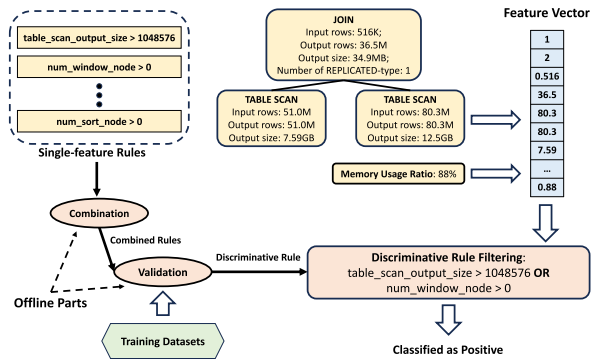

Figure 6: An illustration of discriminative rule filtering.

**Observation 5.** *The learned model can exhibit a high false positive rate (or FP/GT ratio) in certain clusters. Furthermore, confidence scores can serve as a valuable indicator for identifying mispredictions.*

Observation 5 suggests two key principles for improving model performance: (1) A robust mechanism is needed to mitigate excessive false positives by limiting the number of positive predictions made by the learned model. (2) Confidence scores closer to 0 or 1 are more indicative of correct predictions. Building on these insights, we develop a self-tuning quota management mechanism, detailed in Section 4.4.

## 4 SAFELOAD DESIGN
### 4.1 Overview

As shown in Figure 5, SafeLoad operates between the query optimizer and executor, intercepting queries before execution to predict potential memory overload. If a risk is detected, it reroutes the query to a memory-rich serverless cluster, preventing execution failure and resource waste (e.g., CPU time or memory). SafeLoad consists of four key modules that progressively predict overload risks:

**(1) Feature Vector Generation**. This module constructs a multi-dimensional feature vector for each query by combining query plan statistics (e.g., operator cardinalities, node counts) and cluster-level resource metrics (e.g., memory utilization). These features are utilized for both training and inference.

**(2) Discriminative Rule Filtering**. This module uses labeled data from the featurized training data to derive heuristic rules that retain most MO queries while filtering non-MO ones. The rules, combined with logical operators (AND/OR), are applied to eliminate obvious negatives, minimizing overhead and ensuring interpretability.

**(3) Predictive Model with Correction**. As shown in the top right corner of Figure 5, SafeLoad adopts a hybrid architecture, where clusters with insufficient training samples share a global model, while other clusters use their own local models. Additionally, each cluster maintains a vector index for misprediction correction. For queries not filtered by rules, SafeLoad applies cluster-specific or global tree-based models to assess overload risk. To enhance robustness, an in-memory vector index of past false negatives is maintained per cluster. Queries closely resembling previous misclassified MO queries are classified as positive, bypassing the prediction.

**(4) Self-Tuning Quota Management**. SafeLoad dynamically adjusts cluster-specific quotas to mitigate excessive false positives from uncertain predictions. Quota costs, based on prediction entropy and historical false negatives, restrict low-confidence predictions under limited quotas while permitting borderline queries when resources allow, ensuring workload stability.

**Generality and Portability of the SafeLoad Framework**. A central design principle of SafeLoad is its system-agnostic architecture, which enables broad applicability across heterogeneous database engines. The framework deliberately avoids reliance on specific system internals. Instead, SafeLoad's modules are grounded in abstractions that are widely available and transferable. Although optimizer-derived features are employed, these conform to ANSI SQL (ISO/IEC 9075) standards and thus have direct counterparts in both open-source systems (e.g., Presto) and commercial warehouses (e.g., Redshift, Snowflake). Moreover, SafeLoad builds on user-driven workload regularities, such as query repetition patterns, which are independent of specific system internals.

### 4.2 Discriminative Rule Filtering

As noted in Observation 2, MO and non-MO queries can be effectively distinguished based on specific features. For example, if the data scanned by a Table Scan operator is less than 1 MB, the query is unlikely to cause memory overload. SafeLoad leverages such insights to generate the discriminative rule by analyzing positive and negative samples in the training set, as defined below.

Specifically, a *single-feature rule* involves a threshold-based comparison, such as "Table Scan output size > 1 MB", formally `table_scan_output_size > 1048576`. Leveraging Observation 1, three key feature types are identified as highly effective for detecting MO queries: Operator Count, Operator Cardinality, and OOM Indicator (see explanations in Table 2). SafeLoad builds a library of single-feature rules targeting these features, publicly available in [2]. The feature rules and thresholds are derived from extensive business deployment experience. **Discriminative rules** are logical combinations of these single-feature rules, connected using AND/OR operators. In this study, the discriminative rule is specifically formed by combining up to four single-feature rules, as the rule generation process can be completed within one hour, which is an acceptable offline construction time for practical deployment.

The generation of the discriminative rule is depicted in Figure 6. SafeLoad encodes query plan statistics and cluster resource metrics into a feature vector. To ensure robust rule selection, SafeLoad applies the following criteria: (a) *High Positive Retention*: For each feature type, a single-feature rule is chosen that retains the majority of positive samples (e.g., over 95%) while minimizing the retention of negative samples. (b) *High Precision Rule*: An additional rule is selected with a very low negative sample retention rate (e.g., below 3%) while maximizing the retention of positive samples. Logical combinations of these four rules (using AND/OR) are exhaustively evaluated on a validation subset (see Section 5). The final discriminative rule is selected from among the rule combinations with the highest non-MO query filtering rates, while maximizing the retention of MO queries.

Ultimately, the generated discriminative rule is applied online for initial filtering. Queries identified as non-MO are immediately labeled as negative, bypassing further analysis. Potential MO queries that match the rule are forwarded to the predictive model with correction for further evaluation.

### 4.3 Predictive Model with Correction

After rule-based filtering removes easily identifiable non-MO queries, the remaining complex patterns are handled using learned models, which excel at capturing intricate query behaviors [19, 21, 25, 29]. To further enhance detection precision, SafeLoad employs a hybrid architecture combining cluster-specific models and a global model, complemented by a misprediction correction mechanism.

**Base Model for MO Query Detection**. SafeLoad leverages lightweight tree-based models for MO query classification, with XGBoost selected as the default due to its high precision, low inference latency, and proven success in query performance modeling [19, 21, 25, 29]. As demonstrated in Section 6.3, XGBoost consistently outperforms alternatives such as MLP, Random Forest, and LightGBM. The model-agnostic design of this module ensures compatibility with other classifiers if required.

A key challenge is the severe class imbalance in training data [17], with non-MO queries vastly outnumbering MO queries (e.g., a 1:20,276 ratio in SafeBench A2). However, the discriminative rule filtering module mitigates this issue by eliminating a large portion of non-MO queries, substantially improving class balance (e.g., reducing the ratio to 1:464 in SafeBench A2). This preprocessing step not only accelerates training but also enhances the model's ability to distinguish MO queries effectively.

**Hybrid Architecture with Local and Global Models**. To ensure generalization across diverse workloads and cluster configurations, SafeLoad adopts a hybrid architecture (as shown in the top right corner of Figure 5), where some clusters have their own local models, while the remaining clusters share a global model. For clusters with ample labeled data (e.g., over 100 positive samples in the training set), local models are trained to capture cluster-specific patterns. For clusters with fewer positive samples, a global model trained on aggregated data across all clusters is utilized. This design ensures reliable predictions in both data-rich and data-scarce scenarios.

**Misprediction Correction with Vector Search**. As revealed by Observation 4, each MO query is often accompanied by a small number of highly similar counterparts. To leverage this pattern, SafeLoad incorporates a vector search–based mechanism to discover additional true MO queries, thereby reducing the risk of misclassifying MO queries as non-MO. In order to reduce false negatives, SafeLoad integrates FAISS [9] to maintain an in-memory vector index for each database cluster. This index stores feature vectors of MO queries that were misclassified as non-MO during online predictions, enabling corrective actions for future queries. Initially, the vector index is empty and dynamically updates as false negatives are identified. When a query is executed after being misclassified as non-MO by the learned model, and it triggers a memory overload, its feature vector is added to the cluster-specific index (see Figure 7(a)). This allows the system to record and track queries that exhibit overload characteristics.

During future predictions, SafeLoad compares the incoming query's feature vector against the stored vectors in the index using Cosine similarity, with a strict threshold. This threshold, typically as high as 0.9999, ensures that only queries with minor differences from previously overloaded ones are matched, which is particularly relevant in production environments where users often retry failed queries with slight modifications. If a highly similar vector is identified, the system bypasses the learned model and directly classifies the query as MO (see Figure 7(b)). By dynamically adapting to runtime feedback, this mechanism improves detection precision.

### 4.4 Self-tuning Quota Management

SafeLoad incorporates a self-tuning quota management mechanism aimed at adapting to the characteristics of different database clusters to improve prediction precision. This mechanism is especially critical in real-world systems where a high incidence of false positives can result in inefficient resource usage and an erosion of trust in automated decision-making.

**Motivation and Key Scenarios**. This module assigns a quota cost to each incoming query and dynamically adjusts it based on the system's prediction confidence and historical performance. When a query is predicted as positive, it consumes a portion of the cluster's available daily quota. If the remaining quota is insufficient to cover the cost of accepting the query, SafeLoad automatically classifies it as negative. This mechanism is especially valuable in two scenarios. First, under quota scarcity with high-confidence predictions, where most incoming queries are true positives, the system reduces the quota cost proportionally to the cumulative count of recent false negatives. This adjustment enables the system to accept more borderline cases, improving recall and the overall F1-score. Second, in cases of quota surplus with low-confidence predictions, the system increases the cost for predictions with high uncertainty, discouraging acceptance of unreliable queries and reducing the likelihood of false positives. These dynamic adjustments ensure that

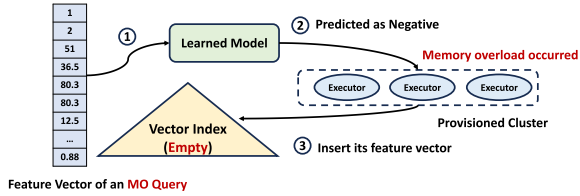 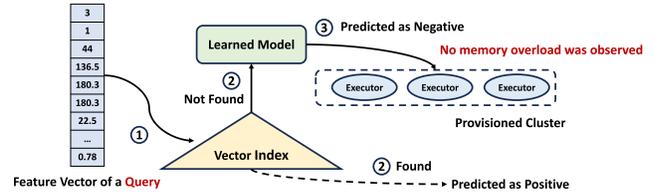

(a) The procedure for the construction of the vector index for each cluster.

(b) The execution workflow of a query involving the Predictive Model with Correction module.

Figure 7: Misprediction correction in SafeLoad.

the system balances precision and recall effectively across diverse clusters.

**Quota Costing Mechanism.** The quota management mechanism in SafeLoad dynamically adjusts the cost of accepting positive predictions to balance resource utilization and prediction reliability.

Specifically, each database cluster $i \in \mathcal{I}$ is allocated an initial daily quota $Q_i^{\text{daily}}$, proportional to the number of MO queries observed on the cluster the previous day. The remaining quota for cluster $i$ at time $t$, denoted as $Q_i(t)$, is initialized as $Q_i(0) = Q_i^{\text{daily}}$ and decreases as queries are accepted.

For an incoming query $s_j$ arriving at time $t_j$, associated with cluster $i$, the learned model assigns a predicted label $\hat{y}_j \in \{0, 1\}$ and a corresponding confidence score $p_j \in [0, 1]$, which indicates the model's confidence that $s_j$ belongs to the positive class.

The decision to accept a positive prediction (i.e., $\hat{y}_j = 1$) is governed by a dynamically computed *quota cost*, denoted $C(s_j)$. This cost is a function of the sample's characteristics and the historical reliability of predictions made for the corresponding cluster. In particular, $C(s_j)$ is determined by the following factors:

(1) <u>Prediction Entropy</u>: The model's confidence score $p_j$ of the sample belonging to the positive class. The uncertainty associated with the prediction is quantified using the binary entropy function:

$$H(p_j) = -p_j \log_2(p_j) - (1 - p_j) \log_2(1 - p_j). \quad (1)$$

Low entropy (near 0) reflects high-confidence predictions (e.g., $p_j$ close to 0 or 1), while higher entropy indicates uncertainty (e.g., $p_j$ near 0.5). Uncertain predictions incur higher quota costs, discouraging acceptance of unreliable decisions. The design of prediction entropy also aligns with Observation 5, which indicates that the closer $p_j$ is to 0 or 1, the more likely the model is to make a correct prediction.

(2) <u>False Negative Count (FNC)</u>: $FNC_i(t)$ tracks the cumulative number of false negatives for cluster $i$ up to time $t$. Higher false negative counts reduce the quota cost to prioritize recall and improve prediction coverage.

The **quota cost** $C(s_j)$ is computed as a weighted combination of these factors, ensuring that the system dynamically adapts to prediction confidence and cluster-specific performance, balancing precision and recall efficiently:

$$C(s_j) = \max\left(1 + \gamma \cdot H(p_j) - \beta \cdot \text{FNC}_i(t), C_{\min}\right). \quad (2)$$

where $\gamma > 0$ is the weight for the prediction uncertainty penalty (entropy $H(p_j)$), $\beta \geq 0$ adjusts for the false negative count ($FNC_i(t)$), and $C_{\min}$ ensures a minimum quota cost to maintain positivity. Under ideal conditions—high model confidence and no false negatives in the cluster, the quota cost defaults to 1.

Table 3: The specification of SafeBench.

| Subset | #(Queries) | #(Clusters) | #(Pos.) | G1 | G2 |
|---|---|---|---|---|---|
| A1 | 52,080,130 | 854 | 4,014 | training | |
| A2 | 50,856,068 | 862 | 2,508 | testing | training |
| A3 | 48,821,677 | 856 | 2,331 | | testing |

A prediction $\hat{y}_j$ is accepted if the remaining quota $Q_i(t_j)$ for cluster $i$ at time $t_j$ is greater than or equal to the cost $C(s_j)$, i.e., Accept $\hat{y}_j = 1$ iff $Q_i(t_j) \geq C(s_j)$. Upon acceptance, the quota is updated as $Q_i(t_j + \Delta t) = Q_i(t_j) - C(s_j)$. If rejected, the quota remains unchanged: $Q_i(t_j + \Delta t) = Q_i(t_j)$.

This mechanism dynamically balances the cost of predictions with model confidence and cluster performance. By penalizing uncertain and error-prone predictions while prioritizing precision when needed, it promotes higher-quality decisions, minimizing false positives and missed positives. This enhances both the operational robustness and overall precision of SafeLoad.

## 5 SAFEBENCH

This section introduces SafeBench, an industrial-grade benchmark open-sourced to advance academic research on preemptively identifying MO queries. To the best of our knowledge, SafeBench is the first benchmark specifically developed for this purpose. The benchmark is publicly available at [2].

**Benchmark Configuration.** Table 3 outlines the specifications of SafeBench, which is constructed from real-world production data collected from Alibaba Cloud's data warehouse system, AnalyticDB [30]. The dataset spans three continuous days, forming three distinct subsets: SafeBench A1 (Day 1), A2 (Day 2), and A3 (Day 3). These subsets include over 150 million analytical queries executed across more than 800 production database clusters. The average CPU time per query is 7.9 seconds (A1), 8.6 seconds (A2), and 8.6 seconds (A3), respectively. To assess the effectiveness of MO query detection methods, we divide the datasets into two evaluation groups. Group G1 uses A1 for training and A2 for testing, while Group G2 uses A2 for training and A3 for testing. By default, SafeBench uses the data from the previous day for training and the data from the current day for testing.

**Labeling of MO Queries.** Ground truth labels were derived from a baseline run of the Table 3 workload without any preventative measures. In this run, a query was marked as MO if its execution logs contained OOM errors or other MO-related failures. We then cross-validated this labeling using runtime traces: if a query's memory demand exceeded available memory at any point during execution, it was confirmed as MO. This two-step procedure ensures accurate identification despite *varying cluster-level conditions*. Although the number of MO queries is small (4014 in A1, 2508 in A2, 2331 in A3), they are disproportionately costly, wasting on average 0.56, 0.59, and 0.51 CPU-hours, respectively. This highlights the importance of accurately detecting these high-impact queries.

**Feature Set**. SafeBench provides detailed profiling for each query from both the query-level and the cluster-level perspectives. Each query is uniquely identified by a query ID with a submission timestamp, and is annotated with metadata such as cluster name, total CPU time, and a binary label indicating whether a memory overload occurred. Each query is further represented by a 163-dimensional feature vector, comprising 147 query-level features and 16 cluster-level features. While derived from AnalyticDB, its query-level features strictly adhere to ANSI SQL abstractions, and cluster-level metrics are limited to fundamental, system-agnostic signals (e.g., CPU load, memory consumption ratios). A categorized summary of the features is provided in Table 2, with detailed definitions available in the open-source SafeBench repository [2].

## 6 EVALUATION

We evaluate SafeLoad in terms of MO query prediction performance and time overheads, covering four evaluation questions:

**EQ1** Prediction Performance (Section 6.2): How does SafeLoad compare to existing baselines in identifying MO queries?

**EQ2** Ablation Study (Section 6.3): What is the contribution of each component in SafeLoad?

**EQ3** Time Overheads (Section 6.4): Are SafeLoad's time overheads competitive and practical for real-world deployment?

**EQ4** Parameter Sensitivity (Section 6.5): Are parameters $\gamma$ and $\beta$ in the quota management robust across a reasonable range, and generalizable across different settings?

### 6.1 Experimental Settings

**Hardware Setup**. All experiments were conducted on a CPU server with Ubuntu 22.04.5, 128 vCPUs (Xeon Platinum 8369B @ 2.70GHz), and 992GiB RAM for large-scale processing and model evaluation.

**Baselines**. Since there is no existing solution specifically designed for identifying MO queries, we construct seven baselines based on relevant prior studies:

(1) **SQL-based-XGB**: Combines TF-IDF feature extraction from SQL statements with XGBoost, inspired by Twitter [21] and CASA [29], two studies on memory consumption prediction.

(2) **Plan-based-XGB**: Trains an XGBoost model using query plan statistics from the optimizer, inspired by Auto-WLM [19], a RedShift query latency estimation framework with applicability to memory consumption estimation.

(3) **MART**: Li et al. [13] proposed MART, which predicts query CPU and I/O time using plan features with a Boosted Regression Tree [5]. We adapt MART for MO query identification.

(4) **LearnedWMP**: LearnedWMP [11] predicts the memory usage of concurrent queries. We use its best-performing variant, LearnedWMP-DNN, as a baseline for MO query identification.

(5) **LS-Rule (Large-Scan-Rule)**: Implements Snowflake's heuristic rules [20], which classify queries as positive when the query plan contains aggregation or filter operators and the table scan size exceeds the daily average.

(6) **Discriminative-Rule**: Directly applies the discriminative rule filtering module from SafeLoad (see Section 4.2).

(7) **Vector-Similarity**: Constructs a vector index from MO query feature vectors in the training set. Queries are classified as positive if their Cosine similarity exceeds 0.9999, matching the threshold used in SafeLoad.

In ablation studies (Section 6.3), the XGBoost model used by SafeLoad for MO query classification is replaced with MLP, LightGBM, and Random Forest, referred to as SafeLoad-MLP, SafeLoad-GBM, and SafeLoad-RF, respectively. Additional experiments further validate the effectiveness of each SafeLoad component. For fairness, query plan statistics used for training and inference in Plan-based-XGB and MART are inherited from the same query-level features included in SafeLoad's feature vector (see Table 2).

**Implementation Details**. For fairness, all baselines involving XGBoost use the same hyperparameter configuration: 500 boosting rounds, a learning rate of 0.05, and a maximum tree depth of 5. For alternative classifiers: (a) the MLP model comprises 4 fully connected layers with hidden dimensions of 256, 128, and 64, using ReLU activations and dropout rates of 0.3 and 0.2, followed by a sigmoid output layer for binary classification; (b) Random Forest is configured with 100 decision trees; (c) LightGBM is trained for 100 boosting rounds with a learning rate of 0.05 and utilizes class-balanced weighting to handle data imbalance. Based on the sensitivity analysis in Section 6.5, parameters in quota management (Equation 2) are defaulted to $\gamma = 1$ and $\beta = 0.5$. The minimum quota cost $C_{\min}$ is fixed at 0.1 to ensure non-negative values.

**Evaluation Protocol**. As outlined in Section 2, MO query detection is framed as a binary classification task, with performance measured using accuracy, precision, recall, and F1-score, prioritizing precision and F1-score as key metrics. We also evaluate CPU time savings for MO queries and the time overheads of online prediction and offline construction. We evaluate SafeLoad and all baselines on both G1 and G2 of the SafeBench benchmark (see Table 3), which supports *deterministic replay*. Each workload trace is replayed on an identically provisioned cluster, ensuring comparable cluster conditions at every query arrival. MO labels are derived once from baseline traces (demand exceeding available memory, cross-checked with OOM logs) and applied consistently across all runs. During evaluation, if a scheduling decision differs (e.g., SafeLoad vs. a baseline), the replay is forked so that only the decision changes while workload and cluster state remain identical. Cluster-level features are also logged as 1-minute snapshots to support TP/FP identification.

### 6.2 Prediction Performance Comparison (EQ1)

We assess SafeLoad's prediction performance compared to baselines from two key perspectives: (1) classification effectiveness using metrics such as precision and F1-score, and (2) efficiency in reducing wasted CPU time across all queries.

**Classification Effectiveness**. Table 4 summarizes the classification results of SafeLoad and baselines on G1 and G2 test datasets, including standard metrics such as precision, recall, and F1-score.

Table 4: Classification performance comparison of SafeLoad and baseline methods on SafeBench G1 and G2 datasets.

| SafeBench G1 | TN | TP | FN | FP | Accuracy | Precision | Recall | F1 |
|---|---|---|---|---|---|---|---|---|
| SQL-based-XGB | 50,852,750 | 174 | 2,334 | 810 | 0.999938 | 0.1768 | 0.0694 | 0.0997 |
| Plan-based-XGB | 50,852,138 | 1,350 | 1,158 | 1,422 | 0.999949 | 0.4870 | 0.5383 | 0.5114 |
| MART | 50,851,233 | 143 | 2,365 | 2,327 | 0.999908 | 0.0579 | 0.0570 | 0.0575 |
| LearnedWMP | 50,849,616 | 941 | 1,567 | 3,944 | 0.999892 | 0.1926 | 0.3751 | 0.2545 |
| Discriminative-Rule | 49,742,527 | 2,393 | 115 | 1,111,033 | 0.978151 | 0.0021 | **0.9541** | 0.0043 |
| LS-Rule | 50,823,588 | 795 | 1,713 | 29,972 | 0.999377 | 0.0258 | 0.3170 | 0.0478 |
| Vector-Similarity | 39,237,558 | 1,692 | 816 | 11,616,002 | 0.771575 | 0.0001 | 0.6746 | 0.0003 |
| SafeLoad | 50,853,048 | 2,203 | 305 | 512 | **0.999984** | **0.8114** | 0.8784 | **0.8436** |
| **SafeBench G2** | TN | TP | FN | FP | Accuracy | Precision | Recall | F1 |
| SQL-based-XGB | 48,816,248 | 7 | 2,324 | 3,098 | 0.999889 | 0.0023 | 0.0030 | 0.0026 |
| Plan-based-XGB | 48,818,272 | 1,175 | 1,156 | 1,074 | 0.999954 | 0.5225 | 0.5041 | 0.5131 |
| MART | 48,816,258 | 1,259 | 1,072 | 3,088 | 0.999915 | 0.2896 | 0.5401 | 0.3771 |
| LearnedWMP | 48,813,028 | 867 | 1,464 | 6,318 | 0.999841 | 0.1206 | 0.3719 | 0.1822 |
| Discriminative-Rule | 47,843,863 | 2,258 | 73 | 975,483 | 0.980018 | 0.0023 | **0.9687** | 0.0046 |
| LS-Rule | 48,788,289 | 641 | 1,690 | 31,057 | 0.999329 | 0.0202 | 0.2750 | 0.0377 |
| Vector-Similarity | 39,337,115 | 1,241 | 1,090 | 9,482,231 | 0.805756 | 0.0001 | 0.5324 | 0.0003 |
| SafeLoad | 48,818,865 | 2,076 | 255 | 481 | **0.999985** | **0.8119** | 0.8906 | **0.8494** |

*SQL-based-XGB* uses TF-IDF to extract keyword feature vectors from SQL statements and applies XGBoost for binary classification. However, it performs poorly on SafeBench G1 and G2, with F1-scores of 0.0997 and 0.0026, respectively, indicating that SQL-only features are insufficient for detecting MO queries, especially without incorporating cluster resource metrics. *Plan-based-XGB* leverages 148 query plan features from the SafeBench dataset and achieves significantly better F1-scores of 0.5114 and 0.5131 on G1 and G2. While this demonstrates the importance of query plan features, its inability to include cluster-level features results in nearly half of MO queries being misclassified.

*MART* and *LearnedWMP* trained boosted regression trees and multilayer perceptrons on plan features to predict query memory consumption. Both methods infer memory overload in our experimental setting by comparing the predicted peak consumption against the available memory. MART attains F1-scores of 0.0575 and 0.3771 on SafeBench G1 and G2, respectively, whereas LearnedWMP achieves 0.2545 and 0.1822. This indirect surrogate decision rule is not tailored to the detection objective and consequently yields poor precision in MO query detection.

*LS-Rule* and *Discriminative-Rule* are lightweight heuristic filters that achieve high accuracy (>0.9) by effectively filtering non-MO queries. However, their F1-scores are extremely low (<0.05), making them unsuitable for production environments.

Notably, *Discriminative-Rule* retains over 93% of MO queries on G1 and G2, effectively filtering non-MO queries. The rule is generated by combining single-feature rules to maximize MO retention while minimizing non-MO retention. Despite being trained separately on G1 and G2, it remains consistent, demonstrating robustness. The final rule, $((($feature_18 > 1 OR feature_63 > 0) AND feature_52 > 1048576 AND feature_161 > 0$))$, maps features to intuitive metrics: `Join` nodes, `TableScan` output size, `Window` nodes, and previous-day cluster OOMs. Its interpretability allows users to optimize SQL queries and avoid MO queries proactively. On G1 and G2 validation sets, positive retention rates reach 93.77% and 95.41%, with negative retention rates of 2.07% and 2.18%, highlighting the rule's effectiveness in preserving MO queries while filtering non-MO queries.

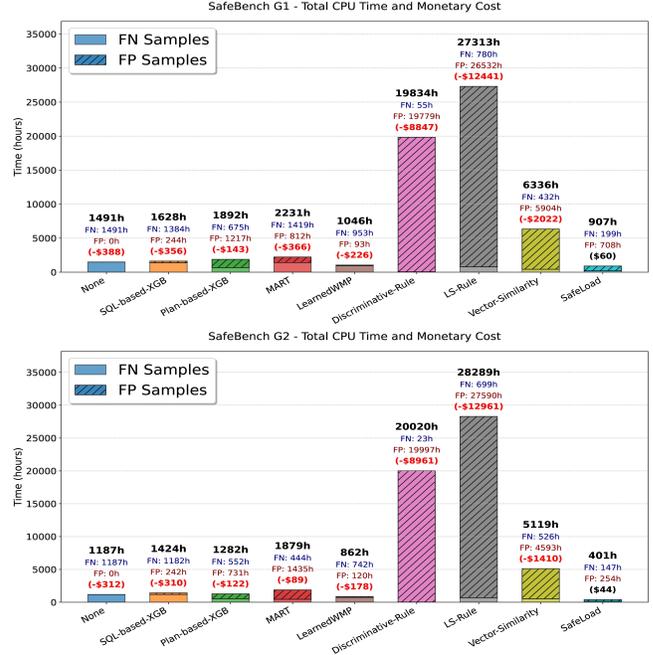

Figure 8: Wasted CPU time and monetary cost across all baseline methods. The monetary cost differential of each baseline relative to SafeLoad (60 USD in G1, 44 USD in G2).

*Vector-Similarity* suffers from low precision due to excessive false positives, as it classifies queries with a Cosine similarity above 0.9999 as positive, often misclassifying non-MO queries in environments with highly similar queries. This underscores the limitations of threshold-based methods and highlights the need for learned models to better capture complex feature distributions and enhance precision.

Overall, SafeLoad achieves SOTA prediction performance, which improves precision by up to 66% over the best baseline and reduces CPU time wasted by MO queries by as much as 8.09× compared to cases without SafeLoad.

**Reduction in Wasted CPU Time and Monetary Cost.** Minimizing CPU time and monetary costs associated with prediction errors is critical in query admission control. Figure 8 shows wasted CPU time and cost for FNs and FPs on G1 and G2. FNs waste CPU on provisioned clusters, while FPs incur overhead on serverless clusters. We adopt an intuitive pricing model: 0.3 USD/CPU-hour for provisioned and 0.5 USD/CPU-hour for serverless clusters, with a complimentary 2,000 CPU-hour quota for MO queries. The OOM queries generated by all instances in our experiments have wasted slightly less than 2000 CPU-hours per day. Consequently, the CPU-hours used by the FP samples, which remain below this threshold, do not incur additional charges to users.

Before deployment, the total CPU waste was 1491.1 and 1187.1 hours on G1 and G2, respectively. SQL-based-XGB slightly reduced this to 1384.1 and 1181.6 hours; Plan-based-XGB achieved greater reductions but still consumed 1217.2 and 730.7 hours on serverless clusters, limiting practicality. Discriminative-Rule, LS-Rule, and Vector-Similarity misclassify many non-MO queries, causing serverless CPU overhead to exceed actual MO CPU usage, making them unsuitable for deployment. In contrast, SafeLoad efficiently identifies MO queries, reducing CPU waste on provisioned clusters to 198.9 and 146.7 hours. Figure 8 shows that deploying SafeLoad has yielded cost savings of 388 USD and 312 USD at the customer level on G1 and G2, respectively.

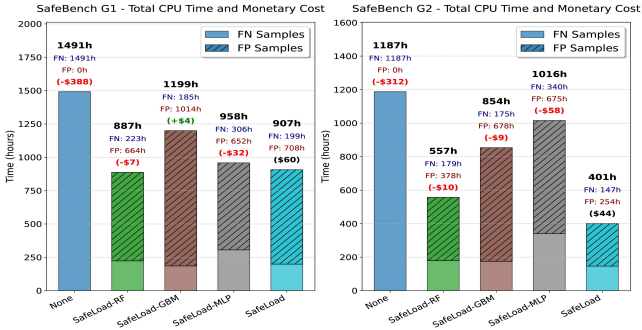

Figure 9: Wasted CPU time and monetary cost for different SafeLoad variants. The monetary cost differential of each variant relative to SafeLoad (60 USD in G1, 44 USD in G2).

Table 5: Ablation study on each SafeLoad component.

| SafeBench G1 | TN | TP | FN | FP | Accuracy | Precision | Recall | F1 |
| --- | --- | --- | --- | --- | --- | --- | --- | --- |
| SafeLoad | 50,853,048 | 2,203 | 305 | 512 | 0.999984 | 0.8114 | 0.8784 | 0.8436 |
| w/o Discriminative Rule Filtering | 50,851,847 | 2,005 | 503 | 1,713 | 0.999956 | 0.5393 | 0.7994 | 0.6441 |
| w/o Global and Local Models | 50,849,660 | 91 | 2,417 | 3,900 | 0.999876 | 0.0228 | 0.0363 | 0.0280 |
| w/o Local Models | 50,852,415 | 2,244 | 264 | 1,145 | 0.999972 | 0.6621 | 0.8947 | 0.7611 |
| w/o Misprediction Correction | 50,853,048 | 2,089 | 419 | 512 | 0.999982 | 0.8032 | 0.8329 | 0.8178 |
| w/o Self-tuning Quota Management | 50,851,702 | 2,286 | 222 | 1,858 | 0.999959 | 0.5516 | 0.9115 | 0.6873 |

| SafeBench G2 | TN | TP | FN | FP | Accuracy | Precision | Recall | F1 |
| --- | --- | --- | --- | --- | --- | --- | --- | --- |
| SafeLoad | 48,818,865 | 2,076 | 255 | 481 | 0.999985 | 0.8119 | 0.8906 | 0.8494 |
| w/o Discriminative Rule Filtering | 48,818,187 | 1,960 | 371 | 1,159 | 0.999969 | 0.6284 | 0.8408 | 0.7193 |
| w/o Global and Local Models | 48,816,940 | 61 | 2,270 | 2,406 | 0.999904 | 0.0247 | 0.0262 | 0.0254 |
| w/o Local Models | 48,818,003 | 2,123 | 208 | 1,343 | 0.999968 | 0.6125 | 0.9108 | 0.7324 |
| w/o Misprediction Correction | 48,818,865 | 1,903 | 428 | 481 | 0.999981 | 0.7982 | 0.8164 | 0.8072 |
| w/o Self-tuning Quota Management | 48,818,559 | 2,097 | 234 | 787 | 0.999979 | 0.7271 | 0.8996 | 0.8042 |

## 6.3 Ablation Studies (EQ2)

*6.3.1 Impact of Key Components.* We performed ablation experiments to assess each component's impact in SafeLoad. Table 5 summarizes the results. Ablation experiments reveal that the **global and local models** contribute the largest performance gain, as their removal causes F1 scores to plummet from above 0.8 to near zero (0.0280 on G1 and 0.0254 on G2), demonstrating their indispensable role in MO query detection. The **discriminative rule filtering** module also yields substantial benefits, with F1 dropping significantly to 0.6441 (G1) and 0.7193 (G2) when removed, alongside increased computational overhead. The **self-tuning quota management** module primarily improves precision, which decreases sharply from 0.8114 to 0.5516 on G1 without it, showing its importance in reducing false positives. The **global model only** setup (without local models) results in moderate declines in F1 (to 0.7611 on G1) and precision (to 0.6621 on G1), indicating the value of local models in capturing cluster-specific patterns. Finally, the **misprediction correction** module produces the smallest but meaningful improvement, with F1 reducing to 0.8178 (G1) when omitted.

Overall, the results demonstrate the critical role of each component in improving detection performance, validating the design choices in our framework.

Table 6: SafeLoad variants with alternative learned models.

| SafeBench G1 | TN | TP | FN | FP | Accuracy | Precision | Recall | F1 |
| --- | --- | --- | --- | --- | --- | --- | --- | --- |
| SafeLoad-RF | 50,853,037 | 2,133 | 375 | 523 | 0.999982 | 0.8031 | 0.8505 | 0.8261 |
| SafeLoad-GBM | 50,852,497 | 2,227 | 281 | 1,063 | 0.999974 | 0.6769 | 0.8880 | 0.7682 |
| SafeLoad-MLP | 50,852,761 | 1,975 | 533 | 799 | 0.999974 | 0.7120 | 0.7875 | 0.7478 |
| SafeLoad | 50,853,048 | 2,203 | 305 | 512 | 0.999984 | 0.8114 | 0.8784 | 0.8436 |

| SafeBench G2 | TN | TP | FN | FP | Accuracy | Precision | Recall | F1 |
| --- | --- | --- | --- | --- | --- | --- | --- | --- |
| SafeLoad-RF | 48,818,828 | 1,987 | 344 | 518 | 0.999982 | 0.7932 | 0.8524 | 0.8218 |
| SafeLoad-GBM | 48,818,486 | 2,057 | 274 | 860 | 0.999977 | 0.7052 | 0.8825 | 0.7839 |
| SafeLoad-MLP | 48,818,855 | 1,704 | 627 | 491 | 0.999977 | 0.7763 | 0.7310 | 0.7530 |
| SafeLoad | 48,818,865 | 2,076 | 255 | 481 | 0.999985 | 0.8119 | 0.8906 | 0.8494 |

*6.3.2 ML Model Variants.* From the perspective of **model prediction performance**, as shown in Table 6, all model variants utilizing the SafeLoad architecture achieve F1-scores above 0.7, significantly outperforming the best baseline, which has an F1-score of 0.5. This highlights the superiority of the SafeLoad architecture in delivering accurate predictions. Among these variants, the XGBoost model in SafeLoad achieves the highest F1-score and precision, underscoring the effectiveness of tree-based models in predicting memory overload.

From the perspective of **reducing CPU time and monetary cost**, as depicted in Figure 9, both SafeLoad-RF and SafeLoad-GBM demonstrate similar performance to SafeLoad in minimizing CPU time waste. However, they fall short in terms of time overhead and prediction precision. For example, SafeLoad-RF incurs an average prediction overhead of 4.2ms per SQL query, whereas SafeLoad achieves a significantly lower overhead of just 0.48ms in scenarios involving the learned model. On the other hand, SafeLoad-GBM has an average prediction precision 0.1 lower than SafeLoad, resulting in a higher number of false positives and additional user overhead. Furthermore, Figure 9 indicates that among all variants of SafeLoad, the configuration leveraging the XGBoost model achieves the highest overall benefit.

## 6.4 Time Overheads (EQ3)

This section analyzes the time overheads of all evaluated methods, divided into online prediction and offline construction overheads.

*6.4.1 Online Prediction Overheads.* The prediction overheads of SQL-based-XGB and Plan-based-XGB are 0.42ms and 0.44ms per query, respectively, about 0.005% of average query latency. The prediction overheads of MART and LearnedWMP are 0.11ms and 0.6ms per query, respectively. Heuristic-rule baselines, Discriminative-Rule, and LS-Rule have negligible overheads of 401ns and 386ns. Vector-Similarity shows low overhead at 0.03ms. SafeLoad processes 97.8% of queries using its rule filtering module with a 401ns overhead. The remaining 2.2%, requiring model prediction and vector search, incur a 0.48ms overhead (~0.006% of query latency). Thus, SafeLoad's online prediction overhead is minimal, making it suitable for production use.

*6.4.2 Offline Construction Overheads.* SQL-based-XGB and Plan-based-XGB incur 485s and 735s of training overhead, respectively. MART and LearnedWMP incur 390s and 15h of training overhead, respectively. LS-Rule has no offline overhead, while Discriminative-Rule requires 34 minutes for rule validation. Vector-Similarity, with fewer MO queries, has a minimal overhead of 0.73ms for vector index construction. SafeLoad's offline process involves XGBoost training, vector index construction, and rule generation. Only 2.2% of samples not filtered by rule filtering are used for model training. On the G1 dataset, vector index construction and model training take 11 seconds, with the total offline overhead at 34 minutes, primarily due to rule generation. Notably, the rules generated for G1 and G2 are identical, suggesting the possibility of reusing rules from the previous day without recomputation. This overhead is reasonable for daily model updates in production.

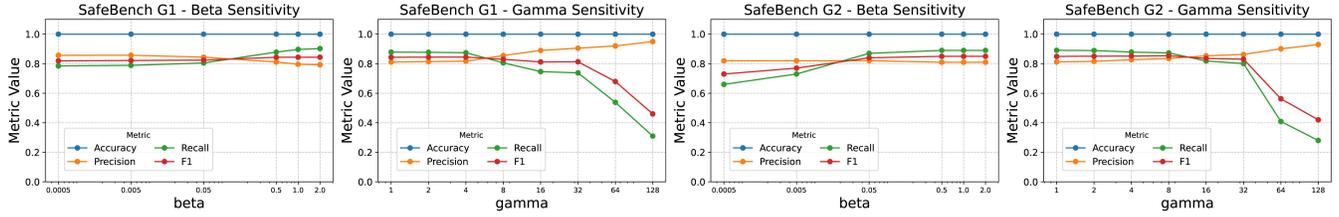

Figure 10: Sensitivity analysis on four classification metrics: $\beta$ on G1, $\gamma$ on G1, $\beta$ on G2, and $\gamma$ on G2 (from left to right).

### 6.5 Parameter Sensitivity Analysis (EQ4)

Through sensitivity analysis, we identified the optimal parameters for SafeLoad's self-tuning quota management: $\gamma = 1$ and $\beta = 0.5$. Figure 10 shows the analysis results for G1 and G2.

For $\beta$ (with $\gamma = 1$), values tested were 0.0005, 0.005, 0.05, 0.5, 1, and 2. Performance remained stable for $\beta$ between 0.5 and 2, with optimal F1 and precision at $\beta = 0.5$. Beyond $\beta = 2$, the quota cost is capped at its minimum, as per Equation 2.

For $\gamma$ (with $\beta = 0.5$), values tested were 1, 2, 4, 8, 16, 32, 64, and 128. Stable performance was observed for $\gamma$ between 1 and 4, but increasing $\gamma$ reduced F1 and precision as fewer positive samples were predicted. This is explained by $\gamma$ modulating model confidence, penalizing low-confidence positives with higher quota costs.

Overall, SafeLoad's classification metrics remain stable across a wide range of parameter values, with $\gamma = 1$ and $\beta = 0.5$ achieving the optimal balance of performance and resource efficiency.

## 7 RELATED WORK

**Query Admission Control**. Existing query admission mechanisms primarily focus on query latency [19, 22, 27], SLA compliance [26], resource isolation [24], and throughput optimization [19]. Q-Cop [22] introduced the concept of a "Query Mix" to predict execution time via linear regression and reject queries likely to exceed time limits. Auto-WLM [19] dynamically adjusted concurrency limits by predicting query length, while Bouncer [27] estimated percentile response times to meet SLO targets. ActiveSLA [26] optimized query admission by evaluating the economic value of SLA fulfillment, and DAC [24] implemented policy-aware resource isolation through transaction classification and dynamic threshold adjustments. However, these methods neglect the critical issue of memory overload. SafeLoad addresses this gap by efficiently identifying MO queries before execution, a novel approach distinct from existing latency and throughput-focused methods.

**Memory Consumption Prediction**. Prior work made progress but left key gaps. Ganapathi et al. [10] used Kernel Canonical Correlation Analysis to predict performance metrics from query plans, focusing primarily on latency with limited attention to memory usage. Li et al. [13] proposed MART, which combines plan features with boosted regression trees to estimate CPU and I/O time; this query resource estimation framework can be adapted for memory prediction. Twitter [21] and CASA [29] utilized SQL text features with XGBoost to predict memory requirements but did not incorporate query plan and cluster-level insights, which are essential for accuracy in dynamic environments [11]. LearnedWMP [18] modeled query templates and memory histograms using MLP and XGBoost, focused on concurrent queries. These approaches struggle to detect memory overload, often overlooking dynamic cluster conditions and peak-demand scenarios. SafeLoad presents a holistic and novel framework that integrates global and local learned models, discriminative rule generation, and self-tuning quota management, enabling proactive identification of MO queries.

**Memory Diagnostics in Industrial Systems**. Snowflake has employed heuristic rules [20] to detect complex queries, but cannot capture intricate patterns. Oracle AWR [3], MySQL Performance Schema [15], BigQuery [4], and AWS Redshift [1] focus on post-execution or runtime analysis, relying on dynamic memory allocation. In contrast, SafeLoad proactively predicts memory overload before execution, combining query features with cluster metrics, setting it apart from existing post-execution methods.

## 8 CONCLUSION

This work presents SafeLoad, an innovative admission control framework that proactively prevents memory overload in cloud data warehouses. By integrating discriminative rule generation, learned models, and a self-tuning quota management mechanism, SafeLoad provides a holistic solution for memory-overloading query detection. Experimental results show a 66% improvement in precision and up to 8.09× reduction in CPU time wastage, highlighting its strong potential to enhance resource management and user experiences in cloud environments.

## ACKNOWLEDGMENTS

This research was funded by the CCF-ApsaraDB Research Fund (Grant No. CCF-Aliyun2024008), the Pioneer R&D Program of Zhejiang (No. 2024C01021), NSFC Grant No. 62402420, NSFC Grant No. U24A20254, and Zhejiang Province "Leading Talent of Technological Innovation Program" (No. 2023R5214). Bo Tang was partially supported by the National Science Foundation of China (NSFC No. 62422206). We thank the anonymous reviewers and the AnalyticDB team for their valuable feedback and suggestions.